\documentstyle[aps]{revtex}

\begin{document}
\title{General relativistic corrections to the Sagnac effect}
\author{A. Tartaglia}
\address{Dip. Fisica, Politecnico, Corso Duca degli Abruzzi 24\\
I-10129 Torino, Italy\\
E-mail: tartaglia@polito.it}
\maketitle

\begin{abstract}
The difference in travel time of corotating and counter-rotating light waves
in the field of a central massive and spinning body is studied. The
corrections to the special relativistic formula are worked out in a Kerr
field. Estimation of numeric values for the Earth and satellites in orbit
around it show that a direct measurement is in the order of concrete
possibilities.
\end{abstract}

\section{ Introduction}

The fact that the round trip time for a light ray moving along a closed path
(thanks to suitably placed mirrors) when its source is on a turntable varies
with the angular speed $\omega $ of the platform may be thought classically
as obvious. Furthermore that time, for a given $\omega $, will be different
if the beam is co-rotating or counter-rotating: longer in the former case,
shorter in the latter. This difference in times, when superimposing the two
oppositely rotating beams, leads to a phase difference with consequent
interference phenomena or, in case of standing waves, to a frequency shift
and ensuing beats. According to Stedman \cite{stedman} this phenomenon was
anticipated by Lodge at the end of the XIX century and by Michelson at the
beginning of the XXth . Experiments were actually performed by Harress \cite
{stedman}\cite{hasselbach}, without being aware of what he observed, and by
Sagnac \cite{sagnac} in 1913 and the interference effect we are speaking of
was since named after him. Sagnac was looking for an ether manifestation and
his approach was entirely classical, but a special relativistic explanation
was soon found giving, to lowest order in $\omega ,$ the same formula for
the time lag between the two light beams 
\begin{equation}
\delta \tau =4\frac S{c^2}\omega  \label{classica}
\end{equation}

$S$ is the area of the projection of the closed path followed by the waves
to contour the platform, orthogonal to the rotation axis; $c$ is the speed
of light and $\omega $ is the rotational velocity of the source/receiver.
The phenomenon is manifested for any kind of waves, including matter waves.
The Sagnac effect has indeed been tested for light, X rays \cite{rostomyan}
and various types of matter waves, such as Cooper pairs \cite{zimmermann},
neutrons \cite{werner}, Ca$^{40}$ atoms \cite{riehle} and electrons \cite
{hasselbach}. A lot of different deductions of (\ref{classica}) have been
given all showing the universal character of the phenomenon; examples are
references \cite{zimmermann} \cite{dresden} \cite{sakurai} \cite{anandan} 
\cite{lefevre} \cite{prade} \cite{hendriks} \cite{scorgie} \cite{dieks} \cite
{mashhoon}. Basically the Sagnac effect is a consequence of the break of the
univocity of simultaneity in rotating systems \cite{rizzi}: this has been
recognized very soon and has also had a direct experimental verification
using identical atomic clocks slowly transported around the world \cite
{allan}.

The Sagnac effect has found a variety of applications both for practical
purposes and fundamental physics, especially after the generalized
introduction, after the 60's, of lasers and ring-lasers \cite{stedman1}
allowing unprecedented precisions in interferometric and frequency shift
measurements. The great accuracy of these measurements poses the problem of
higher order corrections to (\ref{classica}), which have been sought for,
usually in the special relativistic approach. It seems however not to be
unreasonable to consider also general relativistic effects due to the fact
that the ''turntable'' is massive or that the observer is orbiting a massive
and rotating body. This is precisely the scope of the present paper. A
previous work with an aim similar to this was published by Cohen and Mashhoon%
\cite{cohen}; they worked in PPN first order approximation and obtained
results consistent with those presented in this paper.

Section II contains the derivation of the delay in returning to the starting
point for a pair of oppositely rotating light beams in a Kerr field, in the
case of an equatorial trajectory of the rotating observer. Both exact and
approximated results are obtained. In section III the case is treated of a
polar trajectory. Section IV specializes the formulas for a freely falling
observer (circular equatorial orbit). Section V presents some numerical
estimates of the corrections to the usual Sagnac effect, due to the mass and
angular momentum of the Earth. Finally section VI contains a short
discussion of the possibility to measure some of the calculated corrections.

\section{Sagnac effect on a massive rotating body}

The metric describing a rotating black hole (actually a rotating ring
singularity) is the Kerr's one. We begin studying it because it allows for
some exact results and, when suitably approximated, may be used to describe
the gravitational field around a rotating massive body. The Kerr line
element in Boyer-Lindquist space-time coordinates is \cite{gravitation1}: 
\begin{eqnarray*}
ds^{2} &=&\frac{r^{2}-2G\frac{M}{c^{2}}r+\frac{a^{2}}{c^{2}}}{r^{2}+\frac{%
a^{2}}{c^{2}}\cos ^{2}\theta }\left( cdt-\frac{a}{c}\sin ^{2}\theta d\phi
\right) ^{2}- \\
&&\frac{\sin ^{2}\theta }{r^{2}+\frac{a^{2}}{c^{2}}\cos ^{2}\theta }\left[
\left( r^{2}+\frac{a^{2}}{c^{2}}\right) d\phi -adt\right] ^{2}- \\
&&\frac{r^{2}+\frac{a^{2}}{c^{2}}\cos ^{2}\theta }{r^{2}-2G\frac{M}{c^{2}}r+%
\frac{a^{2}}{c^{2}}}dr^{2}-\left( r^{2}+\frac{a^{2}}{c^{2}}\cos ^{2}\theta
\right) d\theta ^{2}
\end{eqnarray*}

Here $M$ is the (asymptotic) mass of the source and $a$ is the ratio between
the angular momentum $J$ and the mass: 
\[
a=\frac JM 
\]

Everything is seen and measured from its effects far away from the black
hole, where space-time is practically flat.

\subsection{Equatorial effect}

Let us now assume that the source/receiver of two oppositely directed light
beams is moving around the rotating black hole which generates the
gravitational field, along a circumference on the equatorial plane. Suitably
placed mirrors send back to their origin both beams after a circular trip
about the central hole.

In this case $r=R=$ constant and $\theta =\pi /2$; the line element is: 
\[
ds^2=\frac{R^2-2G\frac M{c^2}R+\frac{a^2}{c^2}}{R^2}\left( cdt-\frac acd\phi
\right) ^2-\frac 1{R^2}\left[ \left( R^2+\frac{a^2}{c^2}\right) d\phi
-adt\right] ^2 
\]

Let us then assume that the rotation is uniform, so that the rotation angle
of the source/observer is: 
\begin{equation}
\phi _0=\omega _0t  \label{phi0}
\end{equation}

Then 
\begin{equation}
ds^2=\left\{ \frac{R^2-2G\frac M{c^2}R+\frac{a^2}{c^2}}{R^2}\left( 1-\frac a{%
c^2}\omega _0\right) ^2-\frac 1{R^2}\left[ \left( R^2+\frac{a^2}{c^2}\right) 
\frac{\omega _0}c-\frac ac\right] ^2\right\} \left( cdt\right) ^2
\label{ds20}
\end{equation}

For light moving along the same circular path it must be $ds=0$ which
happens when 
\begin{equation}
\frac{R^2-2G\frac M{c^2}R+\frac{a^2}{c^2}}{R^2}\left( 1-\frac a{c^2}\omega
\right) ^2-\frac 1{R^2}\left[ \left( R^2+\frac{a^2}{c^2}\right) \frac \omega %
c-\frac ac\right] ^2=0  \label{luce}
\end{equation}

Now $\omega $ is an unknown; solving (\ref{luce}) for it one finds two
values: 
\begin{equation}
\Omega _{\pm }=\frac 1{\frac{a^2}{c^2}+2G\frac M{c^4R}a^2+R^2}\left( 2G\frac %
M{c^2R}a\pm c\sqrt{\frac{a^2}{c^2}+R^2-2G\frac M{c^2}R}\right)
\label{omega+-}
\end{equation}

$\Omega _{-}$ is actually negative when $R$ exceeds the Schwarzschild limit $%
2G\frac M{c^2}$

The rotation angles for light are then: 
\begin{equation}
\phi _{\pm }=\Omega _{\pm }t  \label{phi+-}
\end{equation}

Eliminating $t$ between (\ref{phi0}) and (\ref{phi+-}): 
\[
\phi _{\pm }=\frac{\Omega _{\pm }}{\omega _0}\phi _0 
\]

Now we proceed applying the geometrical four-dimensional approach that may
be found in \cite{anandan}, \cite{logunov} and \cite{rizzi}. The first
intersection of the world lines of the two light rays with the one of the
orbiting observer after the emission at time $t=0$, is when: 
\begin{eqnarray*}
\phi _{+} &=&\phi _0+2\pi \\
\phi _{-} &=&\phi _0-2\pi
\end{eqnarray*}

i.e. 
\[
\frac{\Omega _{\pm }}\omega \phi _0=\phi _0\pm 2\pi 
\]

Solving for $\phi _0$: 
\begin{equation}
\phi _{0\pm }=\mp \frac{2\pi \omega _0}{\Omega _{\pm }-\omega _0}=\mp \frac{%
2\pi \omega _0}{\frac 1{\frac{a^2}{c^2}+2G\frac M{c^4R}a^2+R^2}\left( 2G%
\frac M{c^2R}a\pm c\sqrt{\frac{a^2}{c^2}+R^2-2G\frac M{c^2}R}\right) -\omega
_0}  \label{lephi}
\end{equation}

The proper time of the rotating observer is deduced from (\ref{ds20})
calling in (\ref{phi0}): 
\[
d\tau =\sqrt{\left( R^2-2G\frac M{c^2}R+\frac{a^2}{c^2}\right) \left( 1-%
\frac a{c^2}\omega _0\right) ^2-\left[ \left( R^2+\frac{a^2}{c^2}\right) 
\frac{\omega _0}c-\frac ac\right] ^2}\frac{d\phi _0}{R\omega _0} 
\]

Finally, integrating between $\phi _{0-}$ and $\phi _{0+}$, we obtain the
Sagnac delay: 
\[
\delta \tau =\sqrt{\left( R^2-2G\frac M{c^2}R+\frac{a^2}{c^2}\right) \left(
1-\frac a{c^2}\omega _0\right) ^2-\left[ \left( R^2+\frac{a^2}{c^2}\right) 
\frac{\omega _0}c-\frac ac\right] ^2}\frac{\phi _{0+}-\phi _{0-}}{R\omega _0}
\]

or explicitly (use \ref{lephi}):

\begin{equation}
\delta \tau =\frac{4\pi }{c^6R}\frac{\left( a^2Rc^2+2GMa^2+R^3c^4\right)
\omega _0-2c^2GMa}{\sqrt{1-\frac 2RG\frac M{c^2}+4G\frac M{c^4R}a\omega
_0-\left( \frac{a^2}{c^4}+2G\frac M{c^6R}a^2+\frac{R^2}{c^2}\right) \omega
_0^2}}  \label{detau}
\end{equation}

This result has some features which are typical of a Kerr geometry. We see
for instance that the delay is zero when the angular speed of the orbiting
observer is 
\[
\omega _n=\frac{2c^2GMa}{a^2Rc^2+2GMa^2+R^3c^4}=2\frac{\frac{GM}{c^2R}\frac a%
{R^2}}{1+2\frac{GM}{c^2R}\frac{a^2}{c^2R^2}+\frac{a^2}{c^2R^2}} 
\]

and provided $a\neq 0$.

This is the velocity of the ''locally non rotating observers'' of the Kerr
geometry \cite{gravitation}: these are equivalent to the static (with
respect to distant stars) observers of the Schwarzschild geometry for which
no Sagnac effect would either be present.

Vice versa when the observer keeps a fixed position with respect to distant
stars ($\omega _0=0$) a time lag, hence a Sagnac effect, is still present,
again under the condition that $a\neq 0$. The time lag is: 
\begin{equation}
\delta \tau _{\left( \omega =0\right) }=\delta \tau _0=-8\pi \frac{GM}{c^4R}%
\frac a{\sqrt{1-2\frac{GM}{c^2R}}}=-8\pi \frac G{c^4R}\frac J{\sqrt{1-2\frac{%
GM}{c^2R}}}  \label{detauo}
\end{equation}

Cohen and Mashhoon\cite{cohen} found the first order approximation of this
same result, which they actually calculated for a static observer sending a
pair of light beams in opposite directions along a closed triangular
circuit, rather than along a circumference.

The delay (\ref{detauo}) is nothing else than the gravitational analog of
the Bohm-Aharonov effect \cite{bohm}. In fact the Sagnac effect is a sort of
inertial Bohm-Aharonov effect \cite{sakurai} \cite{semon} and what we found
is an exact expression for a rotating ring singularity, whereas \cite{harris}
gives an approximated but not simpler result.

Now recalling the Lense-Thirring effect one has a precession velocity \cite
{mashhoon} \cite{stedman1} \cite{soffel} which, in our geometry and
notation, for an equatorial observer, is 
\[
\omega _{LT}=-\frac{GJ}{c^2R^3} 
\]

We see that 
\[
\delta \tau _0=8\frac{\omega _{LT}}{c^2}\frac{\pi R^2}{\sqrt{1-2\frac{GM}{%
c^2R}}} 
\]

The quantity $\delta \tau _0$ doubles the Sagnac delay due to the Lense and
Thirring precession, i.e. to the pure drag by the rotating mass.

\subsection{Approximations}

As we have seen, the deduction of exact results in a Kerr metric, at least
in the special conditions we assumed, is rather straightforward, but of
course in most cases many terms in the equations are very small. This means
that a series of approximations are in order, though it is not necessary to
introduce them from the very beginning as others did \cite{mashhoon1} \cite
{henriksen}.

Let us first assume that $\beta =\omega _0R/c<<1$, consequently developing (%
\ref{detau}) in powers of $\beta $ and retaining only terms up to the second
order; the result is: 
\begin{eqnarray*}
\delta \tau &\simeq &-8\frac \pi {c^4R}GM\frac a{\left( 1-\frac 2RG\frac M{%
c^2}\right) ^{1/2}}+ \\
&&\frac{4\pi R}{c\left( 1-\frac 2RG\frac M{c^2}\right) ^{3/2}}\left( 1+\frac{%
a^2}{R^2c^2}-2\frac{GM}{c^2R}\right) \beta - \\
&&12\pi \frac{GMa}{c^4R}\frac{1+\frac{a^2}{c^2R^2}-\frac 2R\frac{GM}{c^2}}{%
\left( 1-\frac 2RG\frac M{c^2}\right) ^{5/2}}\beta ^2
\end{eqnarray*}

or

\[
\delta \tau \simeq \delta \tau _0+\frac{4\pi }{c\left( 1-\frac 2RG\frac M{c^2%
}\right) ^{3/2}}\left( 1+\frac{a^2}{R^2c^2}-2\frac{GM}{c^2R}\right) \left(
R\beta -\frac{GMa}{c^3R}\frac 3{1-\frac 2RG\frac M{c^2}}\beta ^2\right) 
\]

Now assume also that $\epsilon =\frac{GM}{c^2R}<<1$. To first order in $%
\epsilon $ it is 
\begin{eqnarray*}
\delta \tau &\simeq &-8\frac \pi {c^2}a\epsilon +4\pi \frac Rc\left( 1+\frac{%
a^2}{R^2c^2}\right) \beta + \\
&&\left[ -8\pi \frac Rc+12\pi \frac Rc\left( 1+\frac{a^2}{R^2c^2}\right)
\right] \epsilon \beta - \\
&&12\pi \frac a{c^2}\left( 1+\frac{a^2}{R^2c^2}\right) \epsilon \beta ^2
\end{eqnarray*}

If $\frac a{Rc}$ is at least as small as $\epsilon $:

\[
\delta \tau \simeq -8\frac \pi {c^2}a\epsilon +4\pi \frac Rc\left(
1+\epsilon \right) \beta -12\pi GM\frac a{c^4R}\beta ^2 
\]

Explicitly and calling $\delta \tau _S$ the usual Sagnac effect: 
\begin{eqnarray}
\delta \tau &\simeq &-8\pi a\frac{GM}{c^4R}+4\pi \frac Rc\left( 1+\frac{GM}{%
c^2R}\right) \beta -12\pi \frac{GM}{c^4R}a\beta ^2=  \label{deltatau} \\
&&\delta \tau _S-8\pi a\frac{GM}{c^4R}+4\pi \frac R{c^2}\frac{GM}{c^2}\omega
_0-12\pi R\frac{GM}{c^4}\frac a{c^2}\omega _0^2  \nonumber
\end{eqnarray}

Evidencing the angular momentum: 
\begin{equation}
\delta \tau \simeq \delta \tau _S-8\pi \frac{GJ}{c^4R}+4\pi \frac R{c^2}%
\frac{GM}{c^2}\omega _0-12\pi R\frac{GJ}{c^6}\omega _0^2  \label{deltatau1}
\end{equation}

The usual Sagnac effect is recovered when the terms containing $GM$ and $J$
are negligible. On the other side, a second order correction in $\omega _0^2$
($\beta ^2$) is present only if the angular momentum of the source is
considered.

In these approximations the terms containing $J$ coincide with the first
order (in $J$) corrections to the Schwarzschild field. This fact allows us
to apply the formulas to the simple case of a rotating spherical object
whose radius is $R_{0}.$ Now the angular momentum may be expressed as $%
J=I\Omega _{0}$ where $\Omega _{0}$ is the rotational velocity of the sphere
and $I$ is its moment of inertia. If, just to fix ideas, we assume the
object to have uniform density $\rho ,$ one has: 
\[
I=\frac{8}{15}\rho \pi R_{0}^{5}=\frac{2}{5}MR_{0}^{2} 
\]

Hence the value for $a$ is approximately 
\[
a\simeq \frac{2}{5}R_{0}^{2}\Omega _{0} 
\]

Then for a fixed observer looking at the Earth from the distance $R$ it
comes out 
\[
\delta \tau _0\simeq -\frac{64}{15}\pi ^2\frac{G\rho }{c^4}\frac{R_0^5\Omega
_0}R=-\frac{16}5\pi \frac{GM}{c^4}\frac{R_0^2}R\Omega _0 
\]

\section{Polar (circular) orbit}

It may be interesting to study also a circular trajectory contouring the
central mass passing over the poles. In this case it is again $r=R$, but now 
$\phi =$ const and, retaining uniform motion, $\theta =\omega _0t$; then:

\begin{eqnarray}
ds^2 &=&\frac{R^2-2G\frac M{c^2}R+\frac{a^2}{c^2}}{R^2+\frac{a^2}{c^2}\cos
^2\left( \omega _0t\right) }c^2dt^2-  \label{polare} \\
&&\frac{\sin ^2\left( \omega _0t\right) }{R^2+\frac{a^2}{c^2}\cos ^2\left(
\omega _0t\right) }a^2dt^2-  \nonumber \\
&&\left[ R^2+\frac{a^2}{c^2}\cos ^2\left( \omega _0t\right) \right] \omega
_0^2dt^2  \nonumber
\end{eqnarray}

For light it is of course $ds=0$ which happens when:

\[
\left( R^{2}-2G\frac{M}{c^{2}}R+\frac{a^{2}}{c^{2}}\right) c^{2}-a^{2}\sin
^{2}\theta -\left( R^{2}+\frac{a^{2}}{c^{2}}\cos ^{2}\theta \right)
^{2}\left( \frac{d\theta }{dt}\right) ^{2}=0 
\]

Solving for the angular speed we find that it is no longer constant: 
\[
\frac{d\theta }{dt}=\pm \frac{\sqrt{\left( R^{2}-2G\frac{M}{c^{2}}R+\frac{%
a^{2}}{c^{2}}\right) c^{2}-a^{2}\sin ^{2}\theta }}{R^{2}+\frac{a^{2}}{c^{2}}%
\cos ^{2}\theta } 
\]

This differential equation is easily solvable when $\frac{a^2}{c^2R^2}<<1.$
To first order and assuming $t=0$ when $\theta =0$:

\[
t\simeq \frac{R}{c\left( 1-2G\frac{M}{c^{2}R}\right) ^{1/2}}\theta +\frac{%
a^{2}\left( 1-4G\frac{M}{c^{2}R}\right) }{2c^{3}R\left( 1-2G\frac{M}{c^{2}R}%
\right) ^{3/2}}\int_{0}^{\theta }\cos ^{2}\theta ^{\prime }d\theta ^{\prime
} 
\]

i.e.

\[
t\simeq \frac{R}{c\left( 1-2G\frac{M}{c^{2}R}\right) ^{1/2}}\theta +\frac{%
a^{2}\left( 1-4G\frac{M}{c^{2}R}\right) }{4c^{3}R\left( 1-2G\frac{M}{c^{2}R}%
\right) ^{3/2}}\left( \cos \theta \sin \theta +\theta \right) 
\]

and finally

\[
t\simeq \left[ \frac{R}{c\left( 1-2G\frac{M}{c^{2}R}\right) ^{1/2}}+\frac{%
a^{2}\left( 1-4G\frac{M}{c^{2}R}\right) }{4c^{3}R\left( 1-2G\frac{M}{c^{2}R}%
\right) ^{3/2}}\right] \theta +\frac{a^{2}\left( 1-4G\frac{M}{c^{2}R}\right) 
}{8c^{3}R\left( 1-2G\frac{M}{c^{2}R}\right) ^{3/2}}\sin \left( 2\theta
\right) 
\]

In the same time the rotating observer describes the angle $\theta _0$ while
light travels an angle $2\pi \pm \theta _0$ ($+$ for the co-rotating beam, $%
- $ for the counter-rotating one):

\begin{eqnarray*}
\frac{\theta _{0}}{\omega _{0}} &=&\left[ \frac{R}{c\left( 1-2G\frac{M}{%
c^{2}R}\right) ^{1/2}}+\frac{a^{2}\left( 1-4G\frac{M}{c^{2}R}\right) }{%
4c^{3}R\left( 1-2G\frac{M}{c^{2}R}\right) ^{3/2}}\right] \left( 2\pi \pm
\theta _{0}\right) \pm \\
&&\frac{a^{2}\left( 1-4G\frac{M}{c^{2}R}\right) }{8c^{3}R\left( 1-2G\frac{M}{%
c^{2}R}\right) ^{3/2}}\sin \left( 2\theta _{0}\right)
\end{eqnarray*}

Assume, as we did already, a low speed observer and we expect $2\theta _0$
to be little enough for $\sin \left( 2\theta _0\right) \simeq 2\theta _0$.
Then:

$\frac{\theta _{0}}{\omega _{0}}=\left[ \frac{R}{c\left( 1-2G\frac{M}{c^{2}R}%
\right) ^{1/2}}+\frac{a^{2}\left( 1-4G\frac{M}{c^{2}R}\right) }{%
4c^{3}R\left( 1-2G\frac{M}{c^{2}R}\right) ^{3/2}}\right] \left( 2\pi \pm
\theta _{0}\right) \pm \frac{a^{2}\left( 1-4G\frac{M}{c^{2}R}\right) }{%
4c^{3}R\left( 1-2G\frac{M}{c^{2}R}\right) ^{3/2}}\theta _{0}$

Solving for $\theta _0$ one obtains two results:

\[
\theta _{0\pm }=\pi \frac{2c^2R^2\left( 1-2G\frac M{c^2R}\right) +\frac 12%
a^2\left( 1-\frac{4GM}{c^2R}\right) }{\frac{c^3R}{\omega _0}\left( 1-2G\frac %
M{c^2R}\right) ^{3/2}\mp c^2R^2\left( 1-2G\frac M{c^2R}\right) \mp \frac 12%
a^2\left( 1-\frac{4GM}{c^2R}\right) } 
\]

Finally the difference in round trip times as seen from an inertial
reference frame (recalling the approximation already used for the solution
of this case) results: 
\begin{eqnarray}
t_{+}-t_{-} &=&\frac{\theta _{0+}-\theta _{0-}}{\omega _0}  \label{dtpolo} \\
&\simeq &\pi \frac{R^2}{c^2}\frac{4\left( 1-2\frac{GM}{c^2R}\right) ^2+\frac{%
3+7\beta ^2-6\frac{GM}{c^2R}}{1+\beta ^2-2\frac{GM}{c^2R}}\left( 1-6\frac{GM%
}{c^2R}+8\frac{G^2M^2}{c^4R^2}\right) \frac{a^2}{c^2R^2}}{\left( 1-2\frac{GM%
}{c^2R}\right) ^3+\left( 1-2\frac{GM}{c^2R}\right) ^2\beta ^2}\omega _0 
\nonumber
\end{eqnarray}

For $a=0$ the usual relativistic Sagnac effect is recovered.

To first order in $\epsilon $ (\ref{dtpolo}) becomes:

\[
t_{+}-t_{-}\simeq \pi \frac{R^2}{c^2}\frac{\omega _0}{1+\beta ^2}\left( 4+%
\frac{3+7\beta ^2}{1+\beta ^2}\frac{a^2}{c^2R^2}+\frac 8{1+\beta ^2}\frac{GM%
}{c^2R}\right) 
\]

and finally to first order in $\beta $:

\begin{equation}
t_{+}-t_{-}\simeq \pi \frac{R^2}{c^2}\left( 4+3\frac{a^2}{c^2R^2}+8\frac{GM}{%
c^2R}\right) \omega _0  \label{deti}
\end{equation}

The correction for the moment of inertia of the source is interestingly
independent from $R$; it is indeed: 
\[
3\pi \frac{a^2}{c^4}\omega _0 
\]

which for a sphere in non relativistic approximation is: 
\[
\frac{12}{25}\pi \frac{R_0^4}{c^4}\Omega _0\omega _0 
\]

In order to obtain what the rotating observer sees the result must be
expressed in terms of his proper time; this is done on the base of (\ref
{polare}):

\[
\tau =\int \left\{ \frac{R^2-2G\frac M{c^2}R+\frac{a^2}{c^2}}{R^2+\frac{a^2}{%
c^2}\cos ^2\left( \omega _0t\right) }-\frac{\sin ^2\left( \omega _0t\right) 
}{R^2+\frac{a^2}{c^2}\cos ^2\left( \omega _0t\right) }\frac{a^2}{c^2}-\left[
R^2+\frac{a^2}{c^2}\cos ^2\left( \omega _0t\right) \right] \frac{\omega _0^2%
}{c^2}\right\} ^{1/2}dt 
\]

For short enough time intervals the integrand may be approximated as:

\[
\left[ \frac{1-2G\frac M{c^2R}+\frac{a^2}{c^2R^2}}{1+\frac{a^2}{c^2R^2}}%
-\left( 1+\frac{a^2}{c^2R^2}\right) \frac{R^2\omega _0^2}{c^2}\right]
^{1/2}+O\left( t^2\right) 
\]

and, after integration 
\[
\tau \simeq \left[ \frac{1-2G\frac M{c^2R}+\frac{a^2}{c^2R^2}}{1+\frac{a^2}{%
c^2R^2}}-\left( 1+\frac{a^2}{c^2R^2}\right) \frac{R^2\omega _0^2}{c^2}%
\right] ^{1/2}t 
\]

Adopting the usual approximations: 
\[
\tau \simeq \sqrt{1-2G\frac M{c^2R}-R^2\frac{\omega _0^2}{c^2}}t 
\]

Then 
\[
\delta \tau _p\simeq \sqrt{1-2G\frac M{c^2R}-R^2\frac{\omega _0^2}{c^2}}%
\left( t_{+}-t_{-}\right) 
\]

and explicitly (first order in $\beta $ and $\epsilon $): 
\begin{eqnarray}
\delta \tau _p &\simeq &\pi \frac{R^2}{c^2}\left( 4+3\frac{a^2}{c^2R^2}+4%
\frac{GM}{c^2R}\right) \omega _0=  \label{pollo} \\
&=&\delta \tau _S+\frac \pi {c^4}\left( 3a^2+4RGM\right) \omega _0  \nonumber
\end{eqnarray}

Comparing with the ''equatorial'' situation one has:

\begin{equation}
\delta \tau -\delta \tau _p\simeq -8\pi aG\frac M{c^4R}-3\pi \frac{a^2}{c^4}%
\omega _0  \label{differ}
\end{equation}

\section{Geodesics}

Now we specialize the previous results to a freely falling observer: his
orbit will then be geodesic. If $u^{\mu }$ is the velocity fourvector and $%
\Gamma _{\nu \lambda }^{\mu }$ the Christoffel symbols, the equation of the
geodetics is $\frac{\partial u^{\mu }}{\partial s}+\Gamma _{\nu \alpha
}^{\mu }u^{\alpha }u^{\nu }=0$ where $s$ coincides with the observer's
proper time $\tau $.

Continuing to use Boyer-Lindquist coordinates (generalization of
Schwarzschild coordinates) we are interested in constant radius orbits for
which: 
\begin{eqnarray*}
r &=&R \\
u^{r} &=&0
\end{eqnarray*}

From the geodesic equations and applying these conditions one obtains the
angular speed of the motion about the symmetry axis, $\omega =\frac{u^{\phi }%
}{u^{0}}$; actually there are two different values for the two possible
choices of the rotation with respect to the orientation of the angular
momentum of the source. These angular velocities are in general complicated
functions of $\theta $; this is no problem only when $\theta =$const, i.e. $%
u^{\theta }=0$. Considering this simplified situation and introducing the
Christoffel symbols appropriate to the Kerr metric, the rotation speeds turn
out to be:

\begin{equation}
\omega _{\pm }=\frac{2aGMc^2\pm c^2\sqrt{3a^2G^2M^2+GMc^4R^3}}{a^2GM-c^4R^3}
\label{omegas}
\end{equation}

Recalling now (\ref{detau}) and using (\ref{omegas}) it is possible to find
an exact expression for the time lag for a freely falling object in circular
equatorial orbit.

It is however simpler to develop the (\ref{omegas}) up to first order in $%
\frac a{cR}$: 
\begin{equation}
\omega _{\pm }\simeq \mp \frac cR\sqrt{G\frac M{c^2R}}-\frac 2{R^2}\frac{GM}{%
c^2R}a  \label{ome1}
\end{equation}

Recalling (\ref{deltatau}) and introducing the (\ref{ome1}) we end up with:

\begin{eqnarray*}
\delta \tau _{\pm } &\simeq &8\pi a\frac{GM}{c^{4}R}\pm 4\pi \frac{R}{c}%
\left( 1+\frac{GM}{c^{2}R}\right) \left( \sqrt{\frac{GM}{c^{2}R}}+2\frac{GM}{%
c^{2}R}\frac{a}{cR}\right) \\
&\simeq &\mp 4\pi \frac{R}{c}\sqrt{\frac{GM}{c^{2}R}}+16\pi a\frac{GM}{c^{4}R%
}
\end{eqnarray*}

Now the traditional Sagnac effect is: 
\begin{equation}
\delta \tau _{S\pm }=\pm \frac{4\pi }{c^2}\sqrt{GMR}  \label{libera}
\end{equation}

so we may write 
\begin{equation}
\delta \tau _{\pm }\simeq \delta \tau _{S\pm }+16\pi a\frac{GM}{c^{4}R}
\label{liber1}
\end{equation}

\section{Numerical estimates}

It is interesting to estimate numerical values for the corrections in the
case of the earth as a central body. Now the relevant data are:

\begin{center}
$R_{\oplus }=6.37\times 10^6\quad m$

$\Omega _{\oplus }=7.27\times 10^{-5}\quad rad/s$

$G\frac{M_{\oplus }}{c^2}=4.4\times 10^{-3}\quad m$

$a_{\oplus }=9.81\times 10^8\quad m^2/s$
\end{center}

On the surface of the Earth and if the circular path of the light rays were
the equator, the usual Sagnac delay would be

\begin{equation}
\delta \tau _S=4.12\times 10^{-7}\quad s  \label{tempo}
\end{equation}

This quantity can be converted into a fringe shift multiplying by the
frequency $\nu $ of the light as seen by the observer: 
\begin{equation}
\Delta =\nu \delta \tau _S  \label{delta}
\end{equation}

Considering that for visible light $\nu \sim 10^{14}$ Hz one has a titanic
shift of $\sim 10^7$ fringes. This number makes sense only if the source has
a coherence length as big as at least $123.6$ m which is much but not
impossible. What actually matters, however, is the value of (\ref{delta})
modulo an integer number, which is of course a fraction of a fringe. The
problem is that the knowledge of $\Delta $ requires an accuracy better, say,
than 1 part in $10^8$ and this in turn depends mainly on the accuracy and
stability of the parameters entering the expression of $\delta \tau _S$.

The correction due to the pure mass contribution, $4\pi \frac{R_{\oplus }}{%
c^2}\frac{GM_{\oplus }}{c^2}\Omega _{\oplus }$, is $2.84\times 10^{-16}$ s,
nine orders of magnitude smaller than the main term. The corresponding
fringe shift is $\sim 10^{-2}$.

The correction calling in the moment of inertia of the planet at the lowest
order in $\Omega _{\oplus }$, $-8\pi a\frac{GM}{c^4R}$, is $-1.89\times
10^{-16}$ s. Again a $\sim 10^{-2}$ fringe shift. These shifts are in
principle observable, provided one could find the reference pattern from
which they should be measured, i.e. the value of $\Delta $ modulo an integer
number.

Finally the last correction in (\ref{deltatau}), $-12\pi \frac{GM_{\oplus }}{%
c^6}R_{\oplus }a\Omega _{\oplus }^2$, is $-6.76\times 10^{-28}$:
overwhelmingly small.

Let us now consider an orbiting geodetic observer and assume, just to fix
numbers, that its orbit radius is $R=7\times 10^6$ m.

The main Sagnac term is (\ref{libera}), whose numeric value is: 
\begin{equation}
\delta \tau _S=7.35\times 10^{-6}\quad s  \label{tempo1}
\end{equation}

The fringe shift is $\sim 10^{8}$ and the necessary coherence length would
be greater than $\sim $ $1000$ m. Considering that one is now able to emit
light pulses as short as $\sim 10^{-9}$ s or less, both Sagnac delays (\ref
{tempo}) and (\ref{tempo1}) could be measured directly as such.

The first correction to (\ref{tempo1}) is $16\pi a_{\oplus }\frac{GM_{\oplus
}}{c^4R}$ whose value is $4.16\times 10^{-16}$ s, i.e. $\sim 10^{-2}$
fringes.

If the orbit is polar with the same radius and angular velocity $\omega _0=%
\frac 1R\sqrt{\frac 1RGM}$, the corrections are (see \ref{pollo}) $\frac \pi
{c^4}\left( 3a^2+4RGM\right) \omega _0$, i.e. $\frac \pi {c^4}\frac{%
3a_{\oplus }^2}R\sqrt{G\frac{M_{\oplus }}R}+4\frac \pi {c^4}GM_{\oplus }%
\sqrt{G\frac{M_{\oplus }}R}$. The value of the first term is $1.39\times
10^{-18}$ s ($\sim 10^{-4}$ fringes) and that of the second is $4.84\times
10^{-15}$ s ($\sim 10^{-1}$ fringes). Considering the mass contribution, the
situation is a little bit better than for the equatorial orbit. Furthermore,
when the difference (\ref{differ}) is evaluated we obtain precisely $%
1.39\times 10^{-18}$ s: this, as we said, is of the order of $10^{-4}$
fringes. It is a very small value, but it is obtained comparing two
experimental fringe patterns, without any reference to the basic Sagnac
effect.

\section{Discussion}

Starting from the exact results for a Kerr metric and considering suitable
approximations of them we have obtained the corrections to the Sagnac effect
that the mass and angular momentum of a rotating object introduce. These are
conceptually important, evidencing and strengthening by the way the analogy
between the Sagnac effect and the Bohm-Aharonov effect: particularly
relevant to this purpose is the $\delta \tau _{0}$ of (\ref{detauo}).
Unfortunately, when considering the Earth as the source of the gravitational
field the corrections are indeed very tiny, but per se in the range of what
current optical interference measurements allow, provided a convenient zero
(''pure'' Sagnac term) is experimentally fixed.

When considering devices such as ring lasers, where standing oppositely
propagating waves form, the Sagnac time difference is automatically
converted into a frequency shift and in general a fractional frequency shift
may well be easier to measure than the equivalent fringe shift. Of course
here the difficulty is in stabilizing standing electromagnetic waves around
the Earth, either in space or on the surface of the planet. However what is
hard for light might not be so using radiowaves, provided their Sagnac
effect was not reduced too much.

Apparently there is also the possibility to exploit the difference between
clockwise and counterclockwise rotating observers. In fact, considering (\ref
{libera}) and (\ref{liber1}), we see that: 
\[
\Delta \left( \delta \tau \right) =\delta \tau _{+}-\left| \delta \tau
_{-}\right| =32\pi a\frac{GM}{c^4R} 
\]

Numerically, for satellites orbiting the Earth at $R=7\times 10^{6}$ m, one
has $\Delta \left( \delta \tau \right) =5.8\times 10^{-27}$, corresponding
to a difference in the positions of the interference patterns, of $\sim
10^{-13}$ fringes: absolutely unperceivable.

Summarizing we conclude that experiments to test the existence of the lowest
order general relativistic corrections to the basic Sagnac effect we
computed are in the range of feasibility.

\end{document}